\documentclass[10pt,a4]{article}

\usepackage{amsmath,amsfonts,amssymb,amsthm,ascmac,bm,mathrsfs,pifont,yhmath}
\usepackage{cases,cancel,colortbl,delarray,empheq,feynmp,mathtools,setspace}
\usepackage{graphicx,hhline,multirow,subcaption}
\usepackage{authblk}
\graphicspath{{figures/}} 

\allowdisplaybreaks[4]

\newcommand{\1}{\mbox{1}\hspace{-0.25em}\mbox{l}}
\newcommand{\ket}[1]{\left| #1 \right\rangle}
\newcommand{\bra}[1]{\left\langle #1 \right|}
\newcommand{\bket}[2]{\langle #1 | #2 \rangle}

\begin{document}
\title{Non-Abelian strategies in quantum penny flip game}
\author{Hiroaki Mishima}
\affil{\small Department of Complex Systems Science, Graduate School of Information Science, Nagoya University, Chikusa-ku, Nagoya, Aichi, 464-8601, Japan}
\date{\small January 24, 2018}
\maketitle


\begin{abstract}
In this paper, we formulate and analyze generalizations of the quantum penny flip game.
In the penny flip game, one coin has two states, heads or tails, and two players apply alternating operations on the coin.
In the original Meyer game, the first player is allowed to use quantum (i.e., non-commutative) operations, but the second player is still only allowed to use classical (i.e., commutative) operations.
In our generalized games, both players are allowed to use non-commutative operations, with the second player being partially restricted in what operators they use.
We show that even if the second player is allowed to use ``phase-variable'' operations, which are non-Abelian in general, the first player still has winning strategies.
Furthermore, we show that even when the second player is allowed to choose one from two or more elements of the group $U(2)$, the second player has winning strategies under certain conditions.
These results suggest that there is often a method for restoring the quantum state disturbed by another agent.
\end{abstract}


\maketitle


\section{Introduction}
Game theory is the mathematical language of competitive scenarios in which the outcomes are contingent on the interactions between the strategies of some agents with conflict situations.
The theory of games was originally introduced by von Neumann and Morgenstein~\cite{vNeumann44}, with important contributions by Nash~\cite{Nash}, with the intent of building a theory for predicting economic variation. Since its introduction, the field of game theory has found a diverse range of unforeseen applications including social science, biology, computer science, political science, and, more recently, physics~\cite{Abbott02}.

Games such as the Prisoner's Dilemma~\cite{Poundstone11} and the Hawk--Dove Game~\cite{Maynard82} have proved successful in modeling the evolution of selfish and aggressive behaviors within a population of a species, respectively.
There is now increasing interest in applying game-theoretic techniques in physics.
The models are necessarily idealizations of physical situations.
The situations of interest to game theory are those in which the agents (or players) can select one of a small number of operations (or strategies).
The results of the game, and the corresponding payoffs to the players, are determined through the strategies of all agents.

Quantum computers perform computations by exploiting the quantum mechanical principles of superposition, entanglement, non-locality, and interference~\cite{Eisert04}.
The upsurge of interest in quantum computing has been accompanied with increasing attention in the field of information processing tasks using quantum systems~\cite{Nielsen00}.
At the intersection of game theory and quantum information is the new field of quantum game theory, created in Refs.~\cite{Meyer,Eisert99}.
The two groups independently had the idea of applying the rules of quantum mechanics to game theory.
Replacing the classical probabilities of game theory with quantum amplitudes creates the possibility of new effects resulting from superposition and entanglement.
This occurrence would be a starting point of quantum game theory.
Thus far, quantum game theory has concentrated on detecting these new effects amongst the traditional settings of game theory, but quantum game-theoretic techniques could ultimately be applied in quantum communication~\cite{Brandt98} or quantum computing~\cite{Lee02} protocols.

In a seminal paper~\cite{Meyer}, Meyer introduced the ``quantum penny flip game'' in quantum game theory.
In non-cooperative games, he attempted to apply game theory to quantum mechanics in order to make a thorough investigation of equilibrium behaviors of quantum algorithms.
In the quantum penny flip game that Meyer originated, two players manipulate one invisible coin and try to control the final state of the coin. One player is allowed to use quantum mechanical operations on the coin while another player is allowed to use only classical operations.
Meyer found that strategies that guarantee victory are available to the quantum player.

It has been demonstrated that quantum players are more predominant than classical players \cite{Meyer}.
We here define the meaning of ``predominance'' in a game as follows:
if player {\sf Q} can neutralize another player {\sf P}'s operations while player {\sf P} cannot do so against player {\sf Q}, player {\sf Q} is more predominant than player {\sf P} (see, for example, Sec.~\ref{sec:Meyer_game}). This definition implies that more predominant player {\sf Q} can freely decide the final state of a game if less predominant player {\sf P} has just two operations. Regarding games with malicious rules~\cite{Anand} in which a classical player can win against a full quantum player~\cite{Bang}, this classical player does not have any predominance because he may not be able to lose on purpose, i.e., he cannot freely decide the final state of the game. Being able to win does not constitute predominance. In this paper, we consider that the meaning of ``predominance'' is more strict than that of ``advantage,'' which is often used in other papers.

Although there have been numerous discussions about games with quantum vs. classical players and quantum vs. quantum players, there have been few discussions about games without any ancillary systems with quantum vs. restricted quantum players. 
Such games would be useful in identifying the precise quantum behavior that yields predominance. Namely, the following questions could be answered.
What are the conditions for the existence of the predominance/advantage of a full quantum player under some restrictions of another quantum player?
How much restriction allows predominance/advantage of the full quantum player?

Strategies in the penny flip game can be regarded as a kind of information processing. The quantum penny flip game was introduced to investigate the possible influence of quantum mechanics on information processing~\cite{Meyer,Eisert99,Khan13_qip}.
The purpose of this paper is to investigate whether quantum operations can recover the state of a system disturbed by a classical agent.
Our standpoint can be classified as ``{\it gaming the quantum}''~\cite{Khan13_qic}, which purports to be one of the natural approaches to exploring the quantum landscape for situations that are biasedly or unbiasedly restricted.

The remainder of this paper is organized as follows.
In Sec.~\ref{sec:cqpenny}, we review the classical/quantum penny flip game.
In Sec.~\ref{sec:modi_qpenny}, we change the set of classical player {\sf P}'s commutative operations to various non-commutative ones (see Sec.~\ref{sec:eg1}--\ref{sec:eg4}), find an example of quantum player {\sf Q}'s winning strategy, and calculate the general solutions of player {\sf Q}'s winning strategies.
In Sec.~\ref{sec:con}, we conclude the paper, discuss other research, and mention future work.


\section{Classical/quantum penny flip game}
\label{sec:cqpenny}
Here, we introduce the simple ``penny flip game,'' which is the basis of this paper.


\subsection{Classical version}
The classical penny flip game was introduced by Meyer \cite{Meyer}.
This game has the following rules:
\begin{enumerate}
\renewcommand{\labelenumi}{\roman{enumi})}
  \setlength{\parskip}{0.25truemm}
  \setlength{\itemsep}{0.25truemm}
 \item Players {\sf P} and {\sf Q} have a common penny coin.
 \item The initial state of the penny is heads; the penny is in a box, making it invisible to the players.
 \item Each player can choose whether to flip the penny.
 \item The players can see neither the current state of the penny nor the other player's previous operation.
 \item The sequence of operations is $\text{\sf Q}\to \text{\sf P}\to \text{\sf Q}$.
 \item If the final state is heads, (i.e., the final state is equal to the initial state,) {\sf Q} wins; otherwise, {\sf P} wins.
 \end{enumerate}
The payoff matrix of the game is given in Table~\ref{tb:pmat},
in which $F$, $N$, and $NF$ represent the actions of flip, no flip, and no flip after flip, respectively.
The numbers in the matrix are the payoffs for each player; the first index is for player {\sf P}, and the second index is for player {\sf Q}.
For example, $(-1,1)$ means that player {\sf P} loses and player {\sf Q} wins because the final state is heads.
It is easily verified that the probability of each player winning is $\frac{1}{2}$, and that there exists no pure strategy under Nash equilibrium \cite{Nash}.
The probabilities of the choices of each player are denoted as $\vec{p}:=(p_N,p_F)$ and $
\vec{q}
:=
(q_{NN},q_{NF},q_{FN},q_{FF})$, respectively.
The payoff functions are defined as the expectation of an individual player as $
u_{\sf P}(\vec{p},\vec{q})
=
-u_{\sf Q}(\vec{p},\vec{q})
=
(1-2p_N)[1-2(q_{NF}+q_{FN})]
$.
The mixed-strategy Nash equilibria are given at $
\vec{p}\hspace{0.125em}^*
=\left(\frac{1}{2},\frac{1}{2}\right)
$ and $
\vec{q}\hspace{0.125em}^*
=\left(q_{NN}^*,q_{NF}^*,\frac{1}{2}-q_{NF}^*,\frac{1}{2}-q_{NN}^*\right)
$, where $q_{NF}^*$ and $q_{NN}^*$ may take any value in the range $[0,\frac{1}{2}]$.
Hence, player {\sf P}'s optimal strategy is to choose either $F$ or $N$ with equal probability, and player {\sf Q}'s optimal strategies are to choose either the same or different operations with equal probability.
We find that the average payoffs of both players are zero at the Nash equilibrium.
Altogether, the classical penny flip game is a symmetric, zero-sum, fair game.
\begin{table}[h]
\begin{center}
\caption{Payoff matrix of classical penny flip game.}
\begin{tabular}{cc|c|c|c|c|}
\multicolumn{1}{c}{} &\multicolumn{1}{c}{} &\multicolumn{4}{c}{{\sf Q}}\\
\multicolumn{1}{c}{} &\multicolumn{1}{c}{$({\sf P},{\sf Q})$} & \multicolumn{1}{c}{$NN$} &\multicolumn{1}{c}{$NF$} &\multicolumn{1}{c}{$FN$} &\multicolumn{1}{c}{$FF$} \\ \cline{3-6}
\multirow{2}{*}{{\sf P}} &\multicolumn{1}{c|}{$N$} & $(-1,1)$ & $(1,-1)$ & $(1,-1)$ & $(-1,1)$\\ \cline{3-6}
\multicolumn{1}{c}{} &\multicolumn{1}{c|}{$F$} & $(1,-1)$ & $(-1,1)$ & $(-1,1)$ & $(1,-1)$\\ \cline{3-6}
\end{tabular}
\label{tb:pmat}
\end{center}
\end{table}


\subsection{Quantum version}
\label{sec:Meyer_game}

When discussing unitary quantum operations, we use the following notation: $
\hat{\vec{\sigma}}:=(\hat{\sigma}_1,\hat{\sigma}_2,\hat{\sigma}_3)
$, where $
\hat{\sigma}_1\doteq\scriptsize
\begin{pmatrix}
0&1 \\
1&0
\end{pmatrix},
\hat{\sigma}_2\doteq\scriptsize
\begin{pmatrix}
0&-i \\
i&0
\end{pmatrix}$, and $
\hat{\sigma}_3\doteq\scriptsize
\begin{pmatrix}
1&0 \\
0&-1
\end{pmatrix}
$ are Pauli matrices.

\begin{figure}[htbp]
\begin{center}
\includegraphics[width=86.5mm]{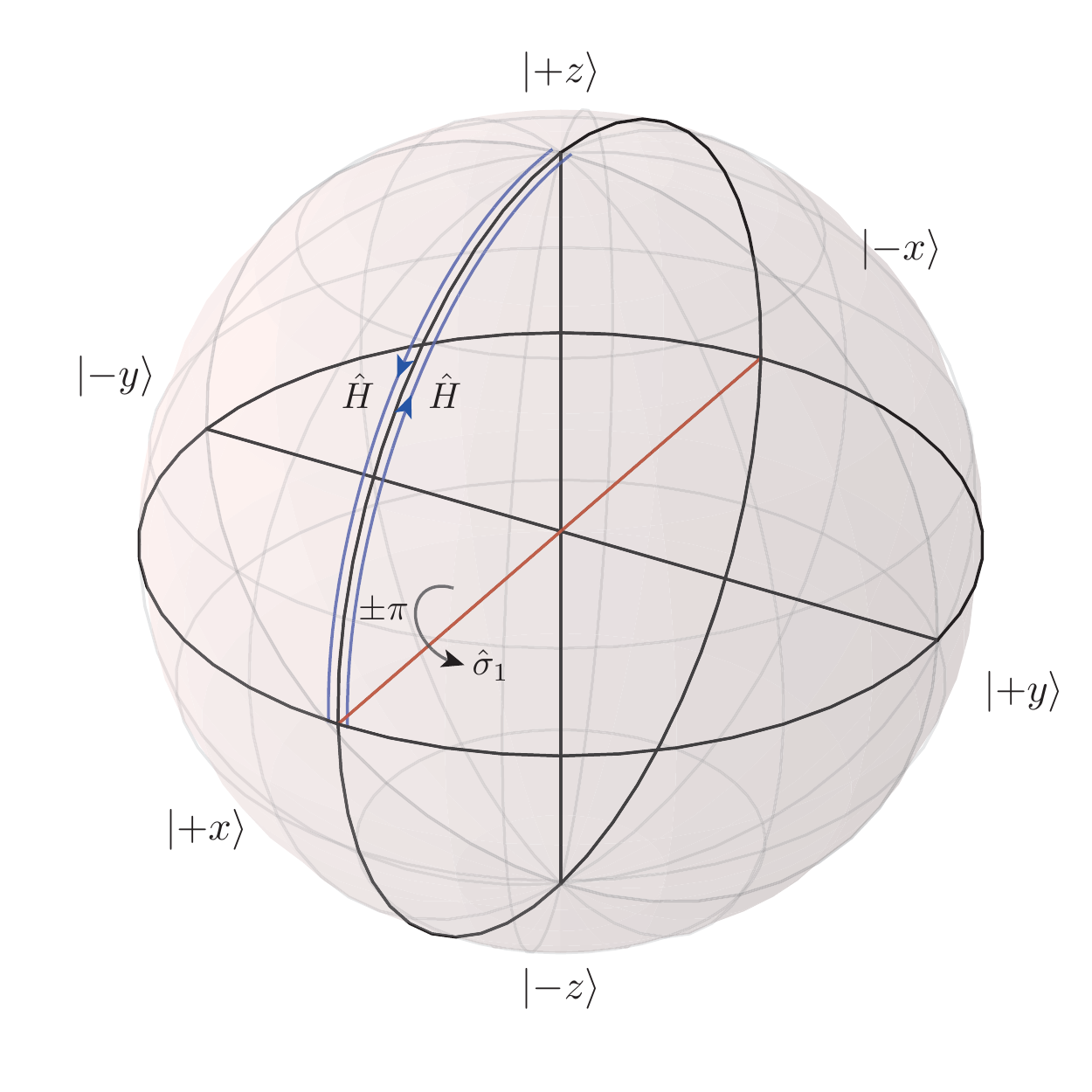}
\label{ps_mu}
\end{center}
\vspace{-2em}
\caption[Text excluding the matrix]{
Winning quantum strategy of Meyer drawn on the Bloch sphere.
Here, we set
$\ket{\pm x}\doteq\frac{1}{\sqrt{2}}
\scriptsize
\begin{pmatrix}
1 \\
\pm 1
\end{pmatrix}$,
$\ket{\pm y}\doteq\frac{1}{\sqrt{2}}
\scriptsize
\begin{pmatrix}
1 \\
\pm i
\end{pmatrix}$,
$\ket{+z}=\ket{0}$
and
$\ket{-z}=\ket{1}$.
The Hadamard transformation, which is the operation of player {\sf Q}, converts the $\ket{+z}$ state to the $\ket{+x}$ state. The operation of $\hat\sigma_1$, which is the coin-flip operation by player {\sf P}, is a rotation around the $x$-axis by $\pi$ radians. Player {\sf P} cannot change the $\ket{+x}$ state by applying the coin-flip operation~[see Eq.~(\ref{eq:qpfg_flow})].
}
\label{fig:Meyer}
\end{figure}

The quantum penny flip game was formulated by Meyer \cite{Meyer}.
In the classical penny flip game, a penny coin takes one of two states: heads or tails.
Meyer introduced a two-state quantum system through the spin of a ``quantum coin.''
In this case, we have to account for quantum properties such as superposition and unitary transformation.
In the quantum penny flip game, only player {\sf Q} can employ a ``quantum strategy.''
Namely, the quantum player can apply arbitrary unitary transformations whereas the classical player can apply only Abelian unitary transformations.
Moving forward, player {\sf P}'s and {\sf Q}'s quantum payoff functions are defined as $\$_{\sf P}=-\$_{\sf Q}=1-2|\bket{f}{i}|^2$, where $\ket{i}$ and $\ket{f}$ are the initial and final states of the coin, respectively.
Meyer showed that player {\sf Q} wins every time if he uses the Hadamard transformation:
\begin{align}
\ket{0}
\xrightarrow[\quad\hat{H}\quad]{\sf Q}
\frac{\ket{0}+\ket{1}}{\sqrt{2}}
\doteq\frac{1}{\sqrt{2}}
\begin{pmatrix}
1 \\
1
\end{pmatrix}
\xrightarrow[\hspace{0.35em}\hat{\sigma}_1\text{ or }{\scriptsize\hat{\1}}\hspace{0.35em}]{\sf P}
\left\{
\begin{array}{ll}
\frac{\ket{1}+\ket{0}}{\sqrt{2}}&\quad\text{if {\sf P} applies $\hat{\sigma}_1$}\\
\frac{\ket{0}+\ket{1}}{\sqrt{2}}&\quad\text{if {\sf P} applies $\hat{\1}$}
\end{array}
\right\}
\xrightarrow[\quad\hat{H}\quad]{\sf Q}
\ket{0},
\label{eq:qpfg_flow}
\end{align}
where $
\ket{0}\doteq\scriptsize
\begin{pmatrix}
1 \\
0
\end{pmatrix}
$ denotes ``heads'' (i.e., spin up), $
\ket{1}\doteq\scriptsize
\begin{pmatrix}
0 \\
1
\end{pmatrix}
$ denotes ``tails'' (i.e., spin down), $
\hat{H}
=
\frac{\hat\sigma_1+\hat\sigma_3}{\sqrt{2}}
\doteq
\frac{1}{\sqrt{2}}\scriptsize
\begin{pmatrix}
1&1 \\
1&-1
\end{pmatrix}$
is the Hadamard transformation, the Pauli matrix $\hat{\sigma}_1$ flips the penny coin, and the identity matrix $\hat{\1}$ leaves the penny coin unchanged.
In the first step, player {\sf Q} applies the Hadamard transformation, $\hat{H}$, which puts the coin into the equal-weight superposition state of heads and tails.
In the second step, player {\sf P} can choose whether to flip the coin, but the superposed state of the coin remains unchanged by either operation selected by player {\sf P}.
In the third step, player {\sf Q} again applies the Hadamard transformation $\hat{H}$, putting the coin back to the initial state because $\hat{H}^2 = \hat{\1}$.
Thus, player {\sf Q} always wins when they open the box.
Hence, in the penny flip game, the quantum strategy is perfectly advantageous against any classical strategy.
It is worth noting that the intermediate state $\ket{+x}$ is a simultaneous eigenstate of player {\sf P}'s operations $\hat{\1}$ and $\hat{\sigma}_1$.
This fact implies that the quantum player {\sf Q} nullifies the operations of player {\sf P}, $\hat{\1}$ or $\hat{\sigma}_1$ (see Fig.~\ref{fig:Meyer}), i.e., player {\sf Q} is predominant.

However, the game proceeds differently if both players are allowed to play with quantum strategies.
Meyer showed that the one-sided advantage is lost in this case (see Theorem 2 of Ref.~\cite{Meyer}).
Although the strategy provided by Meyer is only one of many winning strategies, his example demonstrates the predominance of quantum strategies.
Chappell et al.~\cite{Chappell} provided all of the unitary transformations that are winning strategies for player {\sf Q}:
\begin{align}
\hat{U}^{(1)}_{\sf Q}(\theta,\phi)
=e^{i\delta_1}\exp\left[i\frac{\theta}{2}
\left(a,b\cot\frac{\theta}{2},ab\right)
\cdot\hat{\vec{\sigma}}\right],
\quad
\hat{U}^{(2)}_{\sf Q}(\theta,\phi)
=e^{i\delta_2}e^{i\phi\hat{\sigma}_3/2}\hat{U}^{(1)\dagger}_{\sf Q},
\label{eq:qpfg_wins}
\end{align}
where $\hat{U}^{(1)}_{\sf Q}$ and $\hat{U}^{(2)}_{\sf Q}$ are player {\sf Q}'s first and second operations, respectively, $a=\pm\sqrt{\frac{1}{2}\left(1-\cot^2\frac{\theta}{2}\right)}
$, $
b=\pm 1
$, $
|\theta|\in\left[\frac{\pi}{2},\frac{3\pi}{2}\right]
$, and $
\phi,\delta_1,\delta_2\in[0,2\pi)
$.
By selecting $(\theta,\phi,\delta_1,\delta_2)=(\pi,0,-\frac{\pi}{2},-\frac{\pi}{2})$, the Chappell transformation becomes $\hat{U}^{(1)}_{\sf Q}=\hat{U}^{(2)}_{\sf Q}=\frac{\hat{\sigma}_1+\hat{\sigma}_3}{\sqrt{2}}=\hat{H}$, which corresponds to Meyer's solution.


\section{Modified quantum penny flip games}
\label{sec:modi_qpenny}
In the previous section, we saw that player {\sf Q} can change the state of the coin into a simultaneous eigenstate of the possible operations of player {\sf P} if the operations of {\sf P} are mutually commutative.
This is the winning strategy for player {\sf Q}.
As a direct extension of this observation, we propose a question: if player {\sf P} is allowed to use a restricted class of non-commutative unitary operations, does player {\sf Q} have a winning strategy?
Meyer~\cite{Meyer} showed that if player {\sf P} is also allowed to use any unitary operation, player {\sf Q} has no winning strategies.
Thus, to interpret the question, we must define the class of the operations available to player {\sf P}.


\subsection{Non-Abelian strategy and winning counter-strategy}
\label{sec:eg1}
To begin, we consider a simple modification of the strategy of player {\sf P} by allowing him to use $\hat{\sigma}_3$ instead of $\hat{\1}$ as the non-flipping operation.
Player {\sf P} still uses $\hat{\sigma}_1$ as the flipping operation.
These operators are non-commutative: $[\hat{\sigma}_3,\hat{\sigma}_1]=2i\hat{\sigma}_2\ne 0$; therefore, they generate a non-Abelian group.
In this case, there is no longer a simultaneous eigenstate of player {\sf P}'s operations.
Nevertheless, we found a winning strategy for player {\sf Q}:
\begin{align}
\hat{U}^{(1)}_{\sf Q}\doteq
\frac{1}{\sqrt{2}}
\begin{pmatrix}
1&i\\
i&1
\end{pmatrix},\quad
\hat{U}^{(2)}_{\sf Q}\doteq
\frac{1}{\sqrt{2}}
\begin{pmatrix}
i&-1\\
1&-i
\end{pmatrix}.
\label{eq:eg1_ops}
\end{align}
The game proceeds as follows:
\begin{align}
\ket{0}
&\xrightarrow[\hspace{0.5em}\hat{U}^{(1)}_{\sf Q}\hspace{0.5em}]{\sf Q}
\frac{\ket{0}+i\ket{1}}{\sqrt{2}}
\doteq\frac{1}{\sqrt{2}}
\begin{pmatrix}
1 \\
i
\end{pmatrix}
\xrightarrow[\hspace{0.7em}\hat{\sigma}_1\text{ or }\hat{\sigma}_3\hspace{0.7em}]{\sf P}
\left\{
\begin{array}{ll}
i\frac{\ket{0}-i\ket{1}}{\sqrt{2}}&\text{if {\sf P} applies $\hat{\sigma}_1$}\\
\frac{\ket{0}-i\ket{1}}{\sqrt{2}}&\text{if {\sf P} applies $\hat{\sigma}_3$}
\end{array}
\right\}
\xrightarrow[\hspace{0.5em}\hat{U}^{(2)}_{\sf Q}\hspace{0.5em}]{\sf Q}
\Biggl\{
\begin{array}{ll}
-\ket{0}\\[0.35em]
i\ket{0}
\end{array}
\Biggr..
\label{eq:eg1_flow}
\end{align}
Thus, the final state of the coin is always equivalent to the initial state, heads.
This means that the operations given in Eq.~(\ref{eq:eg1_ops}) constitute a winning strategy for player {\sf Q}.

\begin{figure}[htbp]
\begin{center}
\includegraphics[width=86.5mm]{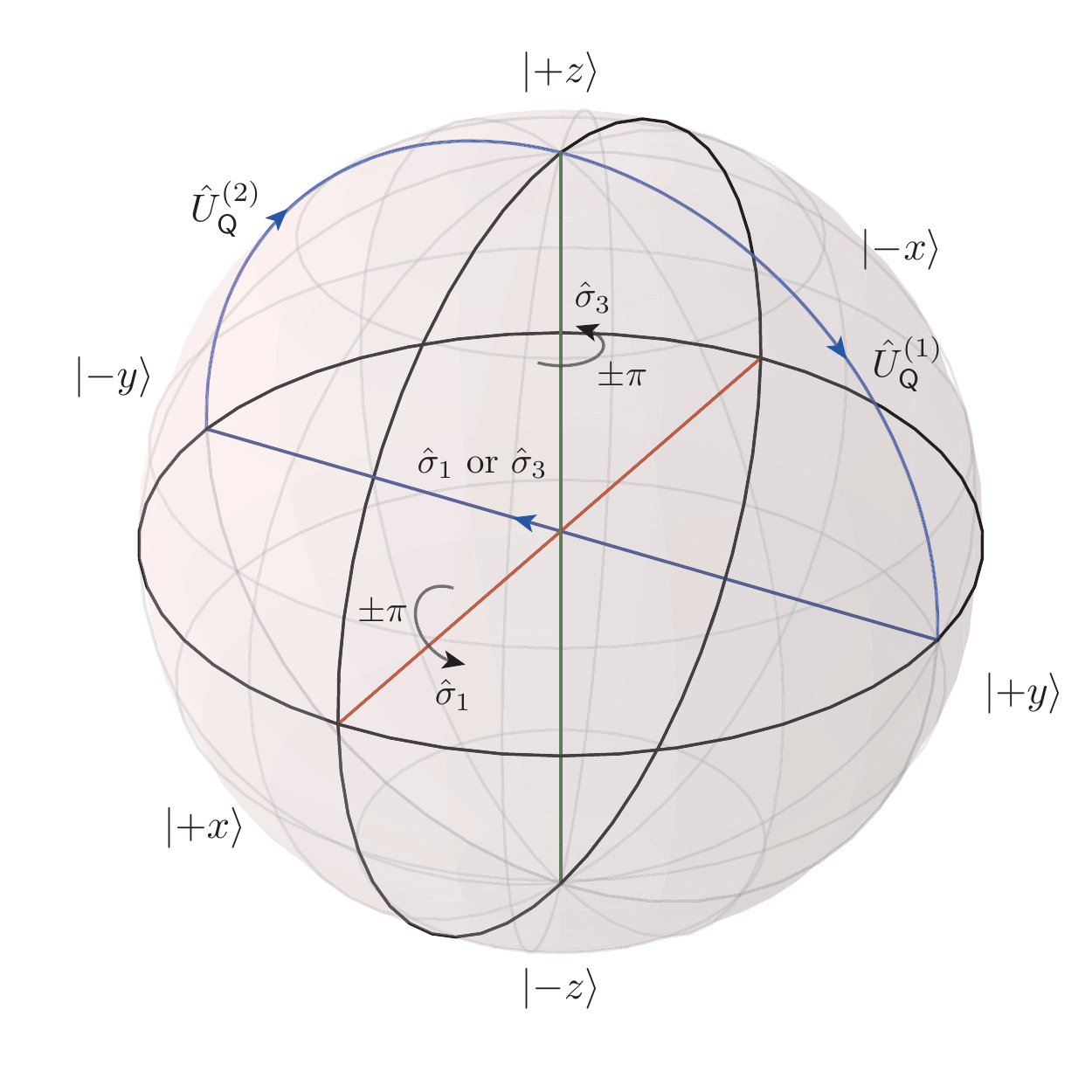}
\label{ps_mu}
\end{center}
\vspace{-2em}
\caption[Text excluding the matrix]{
Winning strategy against operations $\hat{\sigma}_1$ and $\hat{\sigma}_3$ on the Bloch sphere.
Player {\sf P} always converts the state $\ket{+y}$ into $\ket{-y}$~[see Eq.~(\ref{eq:eg1_flow})].
}
\label{fig:our1}
\end{figure}

This strategy utilizes two special states, $\ket{\pm y}\doteq\frac{1}{\sqrt{2}}\scriptsize
\begin{pmatrix}
1\\
\pm i
\end{pmatrix}$,
for which both operations $\hat{\sigma}_1$ and $\hat{\sigma}_3$, i.e., those available to player {\sf P}, have the same effect except for a phase change.
Namely, player {\sf P} must flip the coin through the operations, which implies player {\sf Q} is predominant even if player {\sf P}'s operations are non-commutative.
Thus, player {\sf Q} can always know the state of the coin (see Fig.~\ref{fig:our1}).

Using a method similar to Chappell et al.~\cite{Chappell}, we can obtain all winning strategies for player {\sf Q} in the modified game in which player {\sf P} uses $\hat{\sigma}_1$ and $\hat{\sigma}_3$.
The complete set of the winning strategies are the unitary operators
\begin{align}
\hat{U}^{(1)}_{\sf Q}(\theta,\phi)
=
e^{i\delta_1}\exp\left[i\frac{\theta}{2}
\left(b\cot\frac{\theta}{2},ab,a\right)
\cdot\hat{\vec{\sigma}}\right],
\quad
\hat{U}^{(2)}_{\sf Q}(\theta,\phi)
=e^{i\delta_2}e^{i\phi\hat{\sigma}_3/2}\hat{U}^{(1)\dagger}_{\sf Q}\hat{\sigma}_3,
\label{eq:eg1_wins}
\end{align}
which are parameterized by the same variables as Eq.~(\ref{eq:qpfg_wins}).
By selecting $(\theta,\phi,\delta_1,\delta_2)=(\frac{\pi}{2},0,0,\frac{\pi}{2})$, the general solution in Eq.~(\ref{eq:eg1_wins}) is reduced to Eq.~(\ref{eq:eg1_ops}).


\subsection{Non-Abelian strategy with phase variables and winning counter-strategy}
\label{sec:FN_game}
In this game variant, we introduce a modified flipping operator $\hat{F}$ and a modified non-flipping operator $\hat{N}$ for player {\sf P}:
\begin{align}
\hat{F}(\alpha)
:=
e^{i\alpha\hat{\sigma}_3/2}\hat{\sigma}_1
\doteq
\begin{pmatrix}
0&e^{i\alpha/2}\\
e^{-i\alpha/2}&0
\end{pmatrix},
\quad
\hat{N}(\beta)
:=
e^{i\beta\hat{\sigma}_3/2}
\doteq
\begin{pmatrix}
e^{i\beta/2}&0\\
0&e^{-i\beta/2}
\end{pmatrix},
\label{eq:eg2_ops}
\end{align}
where $\alpha,\beta\in\mathbb{R}$.
In a classical sense, operator $\hat{F}$ flips the coin whereas operator $\hat{N}$ does not, but both introduce phase changes to the quantum state of the coin.
In general, they are non-commutative: $
[\hat{F}(\alpha),\hat{N}(\beta)]
=
2e^{i\alpha\hat{\sigma}_3/2}\hat{\sigma}_2\sin\frac{\beta}{2}
\ne
0$ if $\beta\notin2\pi\mathbb{Z}$.
Using the group composition law of $SU(2)$ \cite{Gellert}, the modified flipping operation can be rewritten as $
i\hat{F}(\alpha)
=
\exp\left[\pm i\frac{\pi}{2}\left(\cos\frac{\alpha}{2},-\sin\frac{\alpha}{2},0\right)\cdot\hat{\vec{\sigma}}\right]
$ 
whose rotation (i.e., flipping) axes are in the same plane of the Bloch sphere.
Even if we replace $\alpha$ with $\alpha+2\pi\mathbb{Z}$, the rotation axis is unchanged.
This is equivalent to the fact that the commutation relation, $
[\hat{F}(\alpha),\hat{F}(\alpha')]
=
2i\hat{\sigma}_3\sin\frac{\alpha-\alpha'}{2}
$, is zero.
We call the operators in Eq.~(\ref{eq:eg2_ops}) a phase-variable strategy, and we call player {\sf P} using this strategy a phase-variable player.
Various operations can be derived from this general case provided that plus--minus signs are arbitrary:
\begin{itemize}
  \setlength{\parskip}{0.25truemm}
  \setlength{\itemsep}{0.25truemm}
 \item By selecting $\alpha,\beta\in 4\pi\mathbb{Z}$, player {\sf P}'s operations become $(\hat{F},\hat{N})=(\hat{\sigma}_1,\hat{\1})$, i.e., those in Meyer's setting.
 \item By selecting $\alpha,\beta\in 2(2\mathbb{Z}+1)\pi$, player {\sf P}'s operations become $(\hat{F},\hat{N})=-(\hat{\sigma}_1,\hat{\1})$, i.e., those in Meyer's setting, except for a sign change.
 \item By selecting $\alpha\in 4\pi\mathbb{Z}$, $\beta\in(4\mathbb{Z}+1)\pi$, player {\sf P}'s operations become $(\hat{F},\hat{N})=(\hat{\sigma}_1,i\hat{\sigma}_3)$, i.e., those defined in Sec.~\ref{sec:eg1} except for a phase change.
 \item By selecting $\alpha\in 2(2\mathbb{Z}+1)\pi$, $\beta\in(4\mathbb{Z}-1)\pi$, player {\sf P}'s operations become $(\hat{F},\hat{N})=-(\hat{\sigma}_1,i\hat{\sigma}_3)$, i.e., those defined in Sec.~\ref{sec:eg1}, except for a phase change.
\end{itemize}
The difference of the phase is not important in these arguments.

We seek a winning strategy for player {\sf Q} against the phase-variable player {\sf P}.
We use the density matrix representation of the coin state to deal with classical and quantum operations on the same footing.
Using the density matrix representation, the game flow is illustrated as:
\begin{align}
\hat{\rho}_0
\xrightarrow[\hspace{0.7em}\hat{U}^{(1)}_{\sf Q}\hspace{0.7em}]{\sf Q}
\hat{\rho}_1
\xrightarrow[\hspace{0.35em}\hat{F}(\alpha)\text{ or }\hat{N}(\beta)\hspace{0.35em}]{\sf P}
\hat{\rho}_2
\xrightarrow[\hspace{0.7em}\hat{U}^{(2)}_{\sf Q}\hspace{0.7em}]{\sf Q}
\hat{\rho}_3.
\label{eq:eg2_flow}
\end{align}
The initial state of the coin is assumed to be heads, $\hat{\rho}_0:=\ket{0}\bra{0}$.
Player {\sf Q} applies a unitary transformation $\hat{U}^{(1)}_{\sf Q}$ on the coin, yielding $
\hat{\rho}_1
:=
{\hat{U}}^{(1)}_{\sf Q}\hat{\rho}_0{{\hat{U}}^{(1)\dagger}_{\sf Q}}
$.
In the next step, player {\sf P} applies the flipping operation $\hat{F}(\alpha)$ with probability $p$ or the non-flipping operation $\hat{N}(\beta)$ with probability $1-p$.
Thus, the density matrix is transformed to $
\hat{\rho}_2
:=
p\hat{F}\hat{\rho_1}\hat{F}^\dagger+(1-p)\hat{N}\hat{\rho}_1\hat{N}^\dagger
$.
The phase parameters $\alpha$ and $\beta$ can be adjusted to yield the strongest strategy for player {\sf P}.
In the final step, player {\sf Q} applies another unitary transformation $\hat{U}^{(2)}_{\sf Q}$, yielding $
\hat{\rho}_3
:=
{\hat{U}}^{(2)}_{\sf Q}\hat{\rho}_2{{\hat{U}}^{(2)\dagger}_{\sf Q}}
$.
Thus, the density matrix of the final state is
\begin{align}
\hat{\rho}_3
=
p{\hat{U}}^{(2)}_{\sf Q}\hat{F}{\hat{U}}^{(1)}_{\sf Q}\hat{\rho}_0{\hat{U}}^{(1)\dagger}_{\sf Q}\hat{F}^\dagger{\hat{U}}^{(2)\dagger}_{\sf Q}+(1-p){\hat{U}}^{(2)}_{\sf Q}\hat{N}{\hat{U}}^{(1)}_{\sf Q}\hat{\rho}_0{\hat{U}}^{(1)\dagger}_{\sf Q}\hat{N}^\dagger{\hat{U}}^{(2)\dagger}_{\sf Q}.
\label{eq:eg2_rho3}
\end{align}
A perfect strategy for player {\sf Q} requires that $\hat{\rho}_3=\hat{\rho}_0$ for arbitrary flip probability $p$; thus, the following equations must hold:
\begin{empheq}[right={\empheqrbrace=\hat{\rho}_0.}]{align}
{\hat{U}}^{(2)}_{\sf Q}\hat{F}{\hat{U}}^{(1)}_{\sf Q}\hat{\rho}_0{\hat{U}}^{(1)\dagger}_{\sf Q}\hat{F}^\dagger{\hat{U}}^{(2)\dagger}_{\sf Q}
\label{eq:eg2_rho3_1}
\\
{\hat{U}}^{(2)}_{\sf Q}\hat{N}{\hat{U}}^{(1)}_{\sf Q}\hat{\rho}_0{\hat{U}}^{(1)\dagger}_{\sf Q}\hat{N}^\dagger{\hat{U}}^{(2)\dagger}_{\sf Q}
\label{eq:eg2_rho3_2}
\end{empheq}
Using $\hat{\rho}_0=\scriptsize\frac{\hat{\1}+\hat{\sigma}_3}{2}$, we can rewrite Eq.~(\ref{eq:eg2_rho3_2}) as $[\hat{U}^{(2)}_{\sf Q}\hat{N}\hat{U}^{(1)}_{\sf Q},\hat{\sigma}_3]=0$.
From this, we can derive a relation between $\hat{U}^{(1)}_{\sf Q}$ and $\hat{U}^{(2)}_{\sf Q}$, i.e., $\hat{U}^{(2)}_{\sf Q}\hat{N}\hat{U}^{(1)}_{\sf Q}=e^{i\delta_2}e^{i\phi\hat{\sigma}_3/2}$, which is equivalent to 
\begin{align}
\hat{U}^{(2)}_{\sf Q}
=
e^{i\delta_2}e^{i\phi\hat{\sigma}_3/2}\hat{U}^{(1)\dagger}_{\sf Q}\hat{N}^\dagger(\beta)
\label{eq:eg2_UQ2}
\end{align}
where $\phi,\delta_2\in[0,2\pi)$.
By substituting Eq.~(\ref{eq:eg2_UQ2}) into Eq.~(\ref{eq:eg2_rho3_1}), we obtain
\begin{align}
e^{i\phi\hat{\sigma}_3/2}\hat{U}^{(1)\dagger}_{\sf Q}\hat{N}^\dagger\hat{F}\hat{U}^{(1)}_{\sf Q}\hat{\rho}_0\hat{U}^{(1)\dagger}_{\sf Q}\hat{F}^\dagger\hat{N}\hat{U}^{(1)}_{\sf Q}e^{-i\phi\hat{\sigma}_3/2}
&=
\hat{\rho}_0.
\label{eq:eg2_rel1_UQ1}
\end{align}
Furthermore, Eq.~(\ref{eq:eg2_rel1_UQ1}) can be rewritten as $
[\hat{U}^{(1)\dagger}_{\sf Q}\hat{N}^\dagger\hat{F}\hat{U}^{(1)}_{\sf Q},\hat{\sigma}_3]
=
0
$, which implies that $
\hat{U}
:=
\hat{U}^{(1)\dagger}_{\sf Q}\hat{N}^\dagger\hat{F}\hat{U}^{(1)}_{\sf Q}
$ is a linear combination of $\hat{\1}$ and $\hat{\sigma}_3$.
Because $\hat{U}$ is an arbitrary unitary transformation, $\hat{U}$ satisfies $\hat{U}^\dagger\hat{U}=\hat{\1}$.
Furthermore, we need to seek $\hat{U}$ satisfying $\hat{U}\ne\pm\hat{\1}$.
We consider the case $
\hat{U}^{(1)\dagger}_{\sf Q}\hat{N}^\dagger\hat{F}\hat{U}^{(1)}_{\sf Q}=\pm\hat{\sigma}_3
$, i.e.,
\begin{align}
\hat{U}^{(1)\dagger}_{\sf Q}\hat{N}^\dagger\hat{F}
=
\pm\hat{\sigma_3}\hat{U}^{(1)\dagger}_{\sf Q}.
\label{eq:eg2_rel2_UQ1}
\end{align}
The unitary operator $\hat{U}^{(1)}_{\sf Q}$ can be parameterized as
\begin{align}
\hat{U}^{(1)}_{\sf Q}
=
e^{i\delta_1}e^{i\theta\vec n\cdot\hat{\vec{\sigma}}/2}
=
e^{i\delta_1}\left(\cos\frac{\theta}{2}\hat{\1}+i\sin\frac{\theta}{2}\vec n\cdot\hat{\vec{\sigma}}\right),
\label{eq:eg2_rel3_UQ1}
\end{align}
with the parameters $\theta\in\mathbb{R}$, $\vec n=(n_1,n_2,n_3)\in S^2\subset\mathbb{R}^3$, and $\delta_1\in[0,2\pi)$.
By substituting Eq.~(\ref{eq:eg2_rel3_UQ1}) into Eq.~(\ref{eq:eg2_rel2_UQ1}), we obtain the relation
\begin{align}
\left[\cos\frac{\theta}{2}\hat{\1}-i\sin\frac{\theta}{2}\left(n_1\hat{\sigma}_1+n_2\hat{\sigma}_2+n_3\hat{\sigma}_3\right)\right]\hat{N}^\dagger\hat{F}
=
b\hat{\sigma}_3\left[\cos\frac{\theta}{2}\hat{\1}-i\sin\frac{\theta}{2}\left(n_1\hat{\sigma}_1+n_2\hat{\sigma}_2+n_3\hat{\sigma}_3\right)\right],
\label{eq:eg2_rel4_UQ1}
\end{align}
where $b=\pm 1$.
We want to find the parameters $\theta$ and $\hat{n}$ satisfying Eq.~(\ref{eq:eg2_rel4_UQ1}).
From Eq.~(\ref{eq:eg2_ops}), we have $
\hat{N}^\dagger\hat{F}
=
\cos\frac{\varDelta}{2}\hat{\sigma}_1-\sin\frac{\varDelta}{2}\hat{\sigma}_2,
$ where $\varDelta:=\alpha-\beta$.
Comparing both sides of Eq.~(\ref{eq:eg2_rel4_UQ1}), this equation is satisfied if
\begin{empheq}[left=\empheqlbrace]{align}
&\left(n_1\cos\frac{\varDelta}{2}-n_2\sin\frac{\varDelta}{2}\right)\sin\frac{\theta}{2}
=
b n_3\sin\frac{\theta}{2},
\label{eq:eg2_rel5a_UQ1}
\\
&\left(n_1\sin\frac{\varDelta}{2}+n_2\cos\frac{\varDelta}{2}\right)\sin\frac{\theta}{2}
=
b \cos\frac{\theta}{2},
\label{eq:eg2_rel5b_UQ1}
\end{empheq}
which implies $\sin\frac{\theta}{2}\ne 0$. From Eqs.~(\ref{eq:eg2_rel5a_UQ1}) and (\ref{eq:eg2_rel5b_UQ1}), we obtain
\begin{align}
n_2
=
b\cot\frac{\theta}{2}\sec\frac{\varDelta}{2}-n_1\tan\frac{\varDelta}{2},\quad
n_3
=
bn_1\sec\frac{\varDelta}{2}-\cot\frac{\theta}{2}\tan\frac{\varDelta}{2}.
\label{eq:eg2_n2n3_with_n1}
\end{align}
By substituting these into the constraint $n_1^2+n_2^2+n_3^2=1$, we obtain
\begin{align}
n_1^2-2bn_1\cot\frac{\theta}{2}\sin\frac{\varDelta}{2}+\frac{1}{2}\left[\cot^2\frac{\theta}{2}\left(1+\sin^2\frac{\varDelta}{2}\right)-\cos^2\frac{\varDelta}{2}\right]
=0,
\end{align}
and hence
\begin{align}
n_1
=
b\cot\frac{\theta}{2}\sin\frac{\varDelta}{2}+a\cos\frac{\varDelta}{2},
\label{eq:eg2_n1}
\end{align}
where $a:=\pm\sqrt{\frac{1}{2}\left(1-\cot^2\frac{\theta}{2}\right)}$ and $\left|\cot\frac{\theta}{2}\right|\le1$ (i.e., $|\theta|\in\left[\frac{\pi}{2},\frac{3\pi}{2}\right]$).
By substituting Eq.~(\ref{eq:eg2_n1}) into Eq.~(\ref{eq:eg2_n2n3_with_n1}), we obtain
\begin{align}
n_2
=
b\cot\frac{\theta}{2}\cos\frac{\varDelta}{2}-a\sin\frac{\varDelta}{2},\quad
n_3
=
ab.
\end{align}
Combining these with Eqs.~(\ref{eq:eg2_UQ2}) and (\ref{eq:eg2_rel3_UQ1}), we obtain the winning strategy for player {\sf Q}:
\begin{align}
\hat{U}^{(1)}_{\sf Q}(\theta,\phi;\alpha,\beta)
&=
e^{i\delta_1}\exp\left[i\frac{\theta}{2}
\left(b\cot\frac{\theta}{2}\sin\frac{\varDelta}{2}+a\cos\frac{\varDelta}{2},\ b\cot\frac{\theta}{2}\cos\frac{\varDelta}{2}-a\sin\frac{\varDelta}{2},\ ab\right)
\cdot\hat{\vec{\sigma}}\right],\\
\hat{U}^{(2)}_{\sf Q}(\theta,\phi;\alpha,\beta)
&=
e^{i\delta_2}\hat{N}(\phi)\hat{U}^{(1)\dagger}_{\sf Q}\hat{N}^\dagger(\beta)
=
e^{i\delta_2}e^{i\phi\hat{\sigma}_3/2}\hat{U}^{(1)\dagger}_{\sf Q}e^{-i\beta\hat{\sigma}_3/2}.
\end{align}
Even if player {\sf P} can change the phase, player {\sf Q} always possesses winning strategies independent of the probability $p$.
Thus, player {\sf Q} is always at least advantageous.


\subsection{Unrestricted strategy and winning counter-strategy}
\label{sec:eg3}

In the above game variants, the operations of player {\sf P} must be coin flipping or non-flipping operations in the classical sense.
Here, we discard this restriction.
Player {\sf P} is allowed to use one of two arbitrary unitary operators $\hat{U}^{(1)}_{\sf P}$ and $\hat{U}^{(2)}_{\sf P}$.
They do not necessarily yield definitive heads or tails states when they act on a coin in the heads state.
Instead, they can yield superposition states of heads and tails.
In this sense, player {\sf P} also becomes a quantum player.
Player {\sf P} applies $\hat{U}^{(1)}_{\sf P}$ to the coin with probability $p$ or $\hat{U}^{(2)}_{\sf P}$ with probability $1-p$.
In this section, we seek a winning strategy for player {\sf Q}. 

Using density matrices, the game flow is illustrated as
\begin{align}
\hat{\rho}_0
\xrightarrow[\hspace{0.7em}\hat{U}^{(1)}_{\sf Q}\hspace{0.7em}]{\sf Q}
\hat{\rho}_1
\xrightarrow[\hspace{0.5em}\{\hat{U}^{(k)}_{\sf P}\}_{k=1,2}\hspace{0.5em}]{\sf P}
\hat{\rho}_2
\xrightarrow[\hspace{0.7em}\hat{U}^{(2)}_{\sf Q}\hspace{0.7em}]{\sf Q}
\hat{\rho}_3.
\end{align}
The final state of the coin is
\begin{align}
\hat{\rho}_3
:=
p\hat{U}^{(2)}_{\sf Q}\hat{U}^{(1)}_{\sf P}\hat{U}^{(1)}_{\sf Q}\hat{\rho}_0{\hat{U}^{(1)\dagger}_{\sf Q}}\hat{U}^{(1)\dagger}_{\sf P}{{\hat{U}}^{(2)\dagger}_{\sf Q}}+(1-p)\hat{U}^{(2)}_{\sf Q}\hat{U}^{(2)}_{\sf P}\hat{U}^{(1)}_{\sf Q}\hat{\rho}_0{\hat{U}^{(1)\dagger}_{\sf Q}}\hat{U}^{(2)\dagger}_{\sf P}{{\hat{U}}^{(2)\dagger}_{\sf Q}}
\label{eq:eg3_rho3}
\end{align}
where $p\in[0,1]$.
We would like to find $\hat{U}^{(1)}_{\sf Q}$ and $\hat{U}^{(2)}_{\sf Q}$ that yield $\hat{\rho}_3=\hat{\rho}_0$ for arbitrary $p$.
Via arguments similar to the previous section, we obtain the equation $
[\hat{U}^{(2)}_{\sf Q}\hat{U}^{(2)}_{\sf P}\hat{U}^{(1)}_{\sf Q},\hat{\sigma}_3]
=
0$.
From this, we have
\begin{align}
\hat{U}^{(2)}_{\sf Q}
=
e^{i\delta_2}e^{i\theta_2\hat{\sigma}_3/2}\hat{U}^{(1)\dagger}_{\sf Q}\hat{U}^{(2)\dagger}_{\sf P},
\label{eq:eg3_UQ2}
\end{align}
where $\delta_2,\theta_2\in[0,2\pi)$.
By substituting Eq.~(\ref{eq:eg3_UQ2}) into Eq.~(\ref{eq:eg3_rho3}), we obtain
\begin{align}
e^{i\theta_2\hat{\sigma}_3/2}\hat{U}^{(1)\dagger}_{\sf Q}\hat{U}^{(2)\dagger}_{\sf P}\hat{U}^{(1)}_{\sf P}\hat{U}^{(1)}_{\sf Q}\hat{\rho}_0{\hat{U}^{(1)\dagger}_{\sf Q}}\hat{U}^{(1)\dagger}_{\sf P}\hat{U}^{(2)}_{\sf P}\hat{U}^{(1)}_{\sf Q}e^{-i\theta_2\hat{\sigma}_3/2}
=
\hat{\rho}_0.
\label{eq:eg3_rel1_UQ1}
\end{align}
Using $\hat{\rho}_0=\scriptsize\frac{\hat{\1}+\hat{\sigma}_3}{2}$, we can rewrite Eq.~(\ref{eq:eg3_rel1_UQ1}) as $
[\hat{U}^{(1)\dagger}_{\sf Q}\hat{U}^{(2)\dagger}_{\sf P}\hat{U}^{(1)}_{\sf P}\hat{U}^{(1)}_{\sf Q},\hat{\sigma}_3]
=
0
$, which implies
\begin{align}
\hat{U}^{(1)\dagger}_{\sf Q}\hat{U}^{(2)\dagger}_{\sf P}\hat{U}^{(1)}_{\sf P}
=
e^{i\delta_3}e^{i\gamma\hat\sigma_3/2}\hat{U}^{(1)\dagger}_{\sf Q},
\label{eq:eg3_rel2_UQ1}
\end{align}
with the parameters $\delta_3,\gamma\in\mathbb R$.
The unitary operators $\hat{U}^{(1)}_{\sf Q}$ and $\hat{U}^{(k)}_{\sf P}$ can be parameterized as
\begin{align}
\hat{U}^{(1)}_{\sf Q}
&=
e^{i\delta_1}e^{i\theta_1\vec n\cdot\hat{\vec{\sigma}}/2}
=
e^{i\delta_1}\left(\cos\frac{\theta_1}{2}\hat{\1}+i\sin\frac{\theta_1}{2}\vec n\cdot\hat{\vec{\sigma}}\right),
\label{eq:eg3_rel3_UQ1}
\\
\hat{U}^{(k)}_{\sf P}
&=
e^{i\xi_k}e^{i\phi_k\vec m_k\cdot\hat{\vec{\sigma}}/2}
=
e^{i\xi_k}\left(\cos\frac{\phi_k}{2}\hat{\1}+i\sin\frac{\phi_k}{2}\vec m_k\cdot\hat{\vec{\sigma}}\right),
\label{eq:eg3_rel1_UPk}
\end{align}
with the parameters $k\,(=1,2)$, $\delta_1,\xi_k,\theta_1,\phi_k\in\mathbb{R}$, $\vec n:=(n_1,n_2,n_3)\in S^2$, and $\vec m_k:=(m_{k1},m_{k2},m_{k3})\in S^2$.
Here, we need to obtain $\hat{U}^{(1)}_{\sf Q}$ satisfying Eq.~(\ref{eq:eg3_rel2_UQ1}).
By using the law of spherical trigonometry~\cite{Gellert}, we can rewrite the left-hand-side of Eq.~(\ref{eq:eg3_rel2_UQ1}) as
\begin{align}
\hat{U}^{(1)\dagger}_{\sf Q}\hat{U}^{(2)\dagger}_{\sf P}\hat{U}^{(1)}_{\sf P}
=
e^{i(\xi_1-\xi_2-\delta_1)}e^{i\varPhi\vec{\mathfrak{M}}\cdot\hat{\vec{\sigma}}/2}
=
e^{i(\xi_1-\xi_2-\delta_1)}\biggl(\cos\frac{\varPhi}{2}\hat{\1}+i\sin\frac{\varPhi}{2}\vec{\mathfrak{M}}\cdot\hat{\vec{\sigma}}\biggr),
\label{eq:eg3_rel4_UQ1}
\end{align}
where
\begin{align}
\cos\frac{\varphi}{2}
&:=
\cos\frac{\phi_1}{2}\cos\frac{\phi_2}{2}+\vec m_1\cdot\vec m_2\sin\frac{\phi_1}{2}\sin\frac{\phi_2}{2},
\label{eq:eg3_def_varphi}\\
\vec M
&:=
\frac{\vec m_1\sin\frac{\phi_1}{2}\cos\frac{\phi_2}{2}-\vec m_2\cos\frac{\phi_1}{2}\sin\frac{\phi_2}{2}-\vec m_1\times\vec m_2\sin\frac{\phi_1}{2}\sin\frac{\phi_2}{2}}{\sin\frac{\varphi}{2}}\in S^2,
\label{eq:eg3_def_unitvec_M}\\
\cos\frac{\varPhi}{2}
&:=
\cos\frac{\theta_1}{2}\cos\frac{\varphi}{2}+\vec M\cdot\vec n\sin\frac{\theta_1}{2}\sin\frac{\varphi}{2},
\label{eq:eg3_def_varPhi}
\\
\vec{\mathfrak{M}}
&:=
\frac{\vec M\sin\frac{\varphi}{2}\cos\frac{\theta_1}{2}-\vec n\cos\frac{\varphi}{2}\sin\frac{\theta_1}{2}-\vec M\times\vec n\sin\frac{\varphi}{2}\sin\frac{\theta_1}{2}}{\sin\frac{\varPhi}{2}}\in S^2.
\label{eq:eg3_def_unitvec_frakM}
\end{align}
Similarly, the right-hand-side of Eq.~(\ref{eq:eg3_rel2_UQ1}) is rewritten as
\begin{align}
e^{i\delta_3}e^{i\gamma\hat\sigma_3/2}\hat{U}^{(1)\dagger}_{\sf Q}
=
e^{i(\delta_3-\delta_1)}\biggl(\cos\frac{\varTheta}{2}\hat{\1}+i\sin\frac{\varTheta}{2}\vec N\cdot\hat{\vec{\sigma}}\biggr),
\label{eq:eg3_rel5_UQ1}
\end{align}
where
\begin{align}
\cos\frac{\varTheta}{2}
&:=
\cos\frac{\gamma}{2}\cos\frac{\theta_1}{2}+n_3\sin\frac{\gamma}{2}\sin\frac{\theta_1}{2},
\label{eq:eg3_def_varTheta}
\\
\vec N
&:=
-\frac{1}{\sin\frac{\varTheta}{2}}
\begin{pmatrix}
(n_1\cos\frac{\gamma}{2}+n_2\sin\frac{\gamma}{2})\sin\frac{\theta_1}{2}\\
(n_2\cos\frac{\gamma}{2}-n_1\sin\frac{\gamma}{2})\sin\frac{\theta_1}{2}\\
n_3\sin\frac{\theta_1}{2}\cos\frac{\gamma}{2}-\cos\frac{\theta_1}{2}\sin\frac{\gamma}{2}
\end{pmatrix}
\in S^2
\label{eq:eg3_def_unitvec_N}
\end{align}
From Eqs.~(\ref{eq:eg3_rel4_UQ1}) and (\ref{eq:eg3_rel5_UQ1}), we obtain the following relation:
\begin{gather}
\cos\frac{\varPhi}{2}\hat{\1}+i\sin\frac{\varPhi}{2}\vec{\mathfrak{M}}\cdot\hat{\vec{\sigma}}
=
e^{i(\delta_3+\xi_2-\xi_1)}\biggl(\cos\frac{\varTheta}{2}\hat{\1}+i\sin\frac{\varTheta}{2}\vec N\cdot\hat{\vec{\sigma}}\biggr).
\label{eq:eg3_rel6_UQ1}
\end{gather}
 Because $\cos\frac{\varPhi}{2},\sin\frac{\varPhi}{2},\cos\frac{\varTheta}{2},\sin\frac{\varTheta}{2}\in\mathbb{R}$, it must be true that $e^{i(\delta_3+\xi_2-\xi_1)}\in\mathbb{R}$.
We choose the value of $\delta_3$ so as to satisfy $\delta_3+\xi_2-\xi_1\in\pi\mathbb{Z}$.
Namely, we find $e^{i(\delta_3+\xi_2-\xi_1)}=:c$, where $c=\pm 1$.
The value of $c$ is decided from the start of the game.
We obtain a system of linear equations:
\begin{empheq}[left=\empheqlbrace]{align}
&\cos\frac{\varPhi}{2}
=
c\cos\frac{\varTheta}{2},
\label{eq:eg3_rel7a_UQ1}
\\
&\vec{\mathfrak{M}}\sin\frac{\varPhi}{2}
=
c\vec N\sin\frac{\varTheta}{2}.
\label{eq:eg3_rel7b_UQ1}
\end{empheq}
Because Eqs.~(\ref{eq:eg3_rel7a_UQ1}) and (\ref{eq:eg3_rel7b_UQ1}) are equivalent to a system of four linear equations with three unknowns $n_1$, $n_2$, and $n_3$, only three of the four equations are mutually independent.
Selecting Eq.~(\ref{eq:eg3_rel7b_UQ1}), we obtain the matrix equation:
\begin{align}
\hat V
\vec n
=
\cos\frac{\theta_1}{2}
\begin{pmatrix}
M_1\sin\frac{\varphi}{2}\\
M_2\sin\frac{\varphi}{2}\\
M_3\sin\frac{\varphi}{2}-c\sin\frac{\gamma}{2}
\end{pmatrix},
\label{eq:eg3_mateq}
\end{align}
where
\begin{align}
\hat{V}
:=
\sin\frac{\theta_1}{2}
\begin{pmatrix}
\cos\frac{\varphi}{2}-c\cos\frac{\gamma}{2} & -\bigl(M_3\sin\frac{\varphi}{2}+c\sin\frac{\gamma}{2}\bigr) & M_2\sin\frac{\varphi}{2}\\
M_3\sin\frac{\varphi}{2}+c\sin\frac{\gamma}{2} & \cos\frac{\varphi}{2}-c\cos\frac{\gamma}{2} & -M_1\sin\frac{\varphi}{2}\\
-M_2\sin\frac{\varphi}{2} & M_1\sin\frac{\varphi}{2} & \cos\frac{\varphi}{2}-c\cos\frac{\gamma}{2}
\end{pmatrix}.
\label{eq:eg3_def_mat_V}
\end{align}
Notably, Eq.~(\ref{eq:eg3_mateq}) can be solved if the inverse matrix $\hat{V}^{-1}$ exists, i.e., if the determinant of matrix $V$ is non-zero.
The determinant of matrix $\hat{V}$, is calculated as
\begin{align}
\det\hat{V}
&=
2c\sin^3\frac{\theta_1}{2}\biggl(\cos\frac{\varphi}{2}-c\cos\frac{\gamma}{2}\biggr)\biggl(M_3\sin\frac{\varphi}{2}\sin\frac{\gamma}{2}-\cos\frac{\varphi}{2}\cos\frac{\gamma}{2}+c\biggr).
\label{eq:eg3_detV}
\end{align}
We find that winning strategies actually exist for player {\sf Q} when player {\sf P} is allowed to use two arbitrary $U(2)$ operations $\hat{U}^{(k)}_{\sf P}$,
\begin{align}
\hat{U}^{(1)}_{\sf Q}
=
e^{i\delta_1}e^{i\theta_1\vec n\cdot\hat{\vec{\sigma}}/2},\quad
\hat{U}^{(2)}_{\sf Q}
=
e^{i\delta_2}e^{i\theta_2\hat{\sigma}_3/2}\hat{U}^{(1)\dagger}_{\sf Q}\hat{U}^{(2)\dagger}_{\sf P},
\label{eq:eg3_win_ops}
\end{align}
where the Bloch vector for player {\sf Q}'s winning strategies is given by
\begin{align}
\vec n
&=
-\frac{\cot\frac{\theta_1}{2}}{M_3\sin\frac{\varphi}{2}\sin\frac{\gamma}{2}-\cos\frac{\varphi}{2}\cos\frac{\gamma}{2}+c}
\begin{pmatrix}
\bigl(M_1\cos\frac{\gamma}{2}-M_2\sin\frac{\gamma}{2}\bigr)\sin\frac{\varphi}{2}\\[2pt]
\bigl(M_1\sin\frac{\gamma}{2}+M_2\cos\frac{\gamma}{2}\bigr)\sin\frac{\varphi}{2}\\[2pt]
M_3\sin\frac{\varphi}{2}\cos\frac{\gamma}{2}+\cos\frac{\varphi}{2}\sin\frac{\gamma}{2}
\end{pmatrix}.
\label{eq:win_n}
\end{align}
Player {\sf Q} should choose parameters $\theta_1$ and $\gamma$ such that Eq.~(\ref{eq:eg3_detV}) is non-zero so as to make Eq.~(\ref{eq:win_n}) converge, with the provision that player {\sf Q} knows the values of $\varphi$ [see Eq.~(\ref{eq:eg3_def_varphi})] and $c=\pm 1$.
When player {\sf Q} operates the strategies in Eq.~(\ref{eq:eg3_win_ops}), player {\sf Q} always wins independent of the probability $p$.
Thus, player {\sf Q} is always at least advantageous.


\subsection {Multiple strategy and winning counter-strategy}
\label{sec:eg4}
Finally, we propose an even more general game.
We allow player {\sf P} to choose one of $\ell$ elements $\{ \hat{U}^{(j)}_{\sf P} \}_{j=1,\cdots,\ell}$ of the group $U(2)$ as his operation.
We call this a multiple strategy.
We seek to evaluate the existence of a winning strategy for player {\sf Q} in this game.
If all the operators given to player {\sf P} are mutually commutative, a simultaneous eigenvector of these operators exists, and this vector is invariant under operations of player {\sf P}.
Hence, in this case, player {\sf Q} always wins by transforming the initial state vector to the simultaneous eigenvector at the first step and transforming it back to the initial state at the final step.

We can also consider when player {\sf P} has one of $\ell$ elements of the group $U(2)$, which are divided into two types of unitary operations.
We allow player {\sf P} to choose $s$ modified flipping operations $\{\hat{F}(\alpha_{k_{\rm F}})\}_{k_{\rm F}=1,\cdots,s}:=\{e^{i\alpha_{k_{\rm F}}\hat{\sigma}_3/2}\hat{\sigma}_1\}_{k_{\rm F}=1,\cdots,s}$ and $\ell-s$ modified non-flipping operations $\{\hat{N}(\beta_{k_{\rm N}})\}_{k_{\rm N}=s+1,\cdots,\ell}:=\{e^{i\beta_{k_{\rm N}}\hat{\sigma}_3/2}\}_{k_{\rm N}=s+1,\cdots,\ell}$.
Player {\sf P} has at least one of each type of unitary operation, i.e., $1\le s\le\ell-1$.

If all of player {\sf P}'s modified flippping operations $\{\hat{F}(\alpha_{k_{\rm F}})\}_{k_{\rm F}}$ are mutually commutative and all of their modified non-flipping operations $\{\hat{N}(\beta_{k_{\rm N}})\}_{k_{\rm N}}$ are equal to identity $\hat \1$, i.e., $\beta_{k_{\rm N}}\in\pi\mathbb Z$ for all $k_{\rm N}$, we can easily deduce that player {\sf Q} always has a complete set of winning strategies because simultaneous eigenstates exist for player {\sf P}, similar to Sec.~ \ref{sec:Meyer_game}.

If all $\{\hat{F}(\alpha_{k_{\rm F}})\}_{k_{\rm F}}$ are mutually commutative and all $\{\hat{N}(\beta_{k_{\rm N}})\}_{k_{\rm N}}$ are not equal to identity $\hat \1$, no winning strategies exist for player {\sf Q} in general.
However, only for $s=1$ or $\ell-1$, winning strategies do exist for player {\sf Q} because examinations such as that in Sec.~\ref{sec:FN_game} are always available.


\section{Conclusions}
\label{sec:con}

Meyer proposed a quantum version of the penny flip game in which player {\sf P} is allowed to use only classical operations, i.e., flipping or non-flipping, on the coin whereas player {\sf Q} is allowed to use any unitary transformation.
Meyer showed that there is a winning strategy for player {\sf Q}; player {\sf Q} always wins by transforming the initial coin state to a superposition state that is the simultaneous eigenvector of the flipping operation $\hat{\sigma}_1$ and the non-flipping operation $\hat{\1}$, and is hence invariant under any operation of player {\sf P}.
Therefore, player {\sf Q} is always predominant.

In this paper, we proposed and analyzed four generalizations of the quantum penny flip game.

In the first generalization, we allow player {\sf P} to use $\hat{\sigma}_1$ and $\hat{\sigma}_3$ as his operations; in contrast to Meyer's game, these operations are non-commutative and do not admit simultaneous eigenvectors.
Even in this game, we found a simple example and a complete set of winning strategies for player {\sf Q}.
After the first winning operation of player {\sf Q}, the two possible operations of player {\sf P} yield equivalent states; therefore, player {\sf Q} can restore the coin state into the same initial state through his second operation.
This scheme is common among all the winning strategies.
Then, player {\sf Q} is always predominant even if player {\sf P}'s operations are non-commutative.

In the second generalization, we allow player {\sf P} to use phase-changing flipping and non-flipping operations.
In this game, we also found a complete set of winning strategies for player {\sf Q}, with the provision that player {\sf Q} knows the values of the parameters $\alpha$ and $\beta$ in player {\sf P}'s operations.
Thus, player {\sf Q} is always at least advantageous.

In the third generalization, we allow player {\sf P} to use two arbitrary unitary operations.
Even in this game, player {\sf Q} has a set of winning strategies with a suitable choice of parameters.
This fact implies that non-commutativity, phase, and the number of generators of unitary operations are completely unrelated to the existence of winning strategies.
Thus, player {\sf Q} is always at least advantageous.

In the fourth generalization, we allow player {\sf P} to use $\ell\ge 3$ elements of phase-changing flipping and non-flipping operations.
Even in this game, player {\sf Q} has a set of winning strategies if some conditions are satisfied.
Meyer's original game and our first, second, and third generalizations are special cases of this fourth generalized game.
Consequently, we found that even if player {\sf P} has non-Abelian mixed strategies, there were cases in which player {\sf Q} has a set of winning strategies. 

In these games, the purpose of player {\sf Q} was to restore the initial state at the end whereas the purpose of player {\sf P} was to change the coin from the initial state.
In this context, a winning strategy for player {\sf Q} is equivalent to restoration of the initial state against player {\sf P}.
Furthermore, the conditions for the existence of winning strategies were similar to the classification of interference such that the initial state can be always restored.

We hope that the present work provides a new perspective on other quantum games in various fields such as finance~\cite{Hanauske10}.
In the quantum Prisoner's Dilemma~\cite{Eisert99}, the quantum Hawk--Dove game~\cite{Hanauske10}, and the quantum stag hunt game~\cite{Toyota03}, replacing a classical player's operations with a restricted set of quantum operations could change properties of the game.
Especially in the quantum penny flip game~\cite{Meyer} and our modified games, the goals of the two players are to either save the initial state or disturb it.
To guarantee victory, player {\sf Q} needs to set a suitable intermediate state.
This situation can also be represented as quantum information processing, i.e., player {\sf Q} can be regarded as the sender/receiver of information and player {\sf P} can be regarded as an eavesdropper.


\section*{Acknowledgments}

The author is grateful to Tomotoshi Nishino for guiding the research of the present subject, to Shogo Tanimura for careful reading of the manuscript and helpful suggestions for improving it, and to Ryosuke Ishiwata for numerous useful discussions.



\end{document}